\documentclass[a4paper,14pt]{article}
\usepackage[english]{babel}
\usepackage{indentfirst}
\usepackage{epsfig}
\usepackage{mathtools}
\usepackage{authblk}
\usepackage{geometry}
\usepackage{natbib}
\title{Mass-Loss Rates of ``Hot-Jupiter'' Exoplanets with Various Types of Gaseous Envelopes}

\author[1]{A. A. Cherenkov}
\author[1]{D. V. Bisikalo\footnote{\texttt{bisikalo{@}inasan.ru}}}
\author[1]{P. V. Kaigorodov}

\affil[1]{Institute of Astronomy, Russian Academy of Sciences, Moscow, Russia}

\date{\vspace{-8ex}}

\newcommand\Lp[1]{$\mathrm{L_#1}$}

\def\gs{$\mathrm{g\,s}^{-1}$}
\def\gscm{$\mathrm{g\,s^{-1}\,cm^{-2}}$}

\def\Rpl{\ensuremath{R_{\rm pl}}}
\def\Mpl{\ensuremath{M_{\rm pl}}}

\begin{document}
\maketitle

\begin{abstract}

According to the compuations results obtained by Bisikalo et al.~(2013b) for the gas-dynamical effect of stellar winds on exoplanet atmospheres, three types of gaseous envelopes can form around hot Jupiters: closed, quasi-closed, and open. The type of envelope that forms depends on the position of the frontal collision point (where the dynamical pressure of the wind is equal to the pressure of the surrounding atmosphere) relative to the Roche-lobe boundaries. Closed envelopes are formed around planets whose atmospheres lie completely within their Roche lobes. If the frontal collision point is located outside the Roche lobe, the atmospheric material begins to flow out through the Lagrangian points \Lp1 and \Lp2, which can result in the formation of quasi-closed (if the dynamical pressure of the stellar wind stops the outflow through \Lp1) or open gaseous envelopes. The example of the typical hot Jupiter HD\,209458\,b is considered for four sets of atmospheric parameters, to determine the mass-loss rates for the different types of envelopes arising with these parameters. The mass-loss rates based on the modeling results were estimated to be $\dot{M} \leq 10^{9}$\,\gs\ for a closed atmosphere, $\dot{M} \simeq 3 \times 10^{9}$\,\gs\ for a quasi-closed atmosphere, and $\dot{M} \simeq 3 \times 10^{10}$\,\gs\,for an open atmosphere. The matter in the closed and quasi-closed atmospheres flows out mainly through \Lp2, and the matter in open envelopes primarily through \Lp1.

\end{abstract}

\section{Introduction}

``Hot Jupiters'' --- exoplanets with masses comparable to Jupiter's mass but with orbital semi-major axes no greater than 0.1\,AU --- have a number of unique properties due to their proximity to their parent stars. Some of these properties, such as an increase in their atmospheric temperatures due to intense heating by the stellar radiation, are obvious. However, others do not have such a clear origin, as they are associated with a more complicated series of physical processes. For example, proximity to the star may lead to the outflow of some matter from the planetary atmosphere toward the star. Moreover, a small distance to the star results in a high orbital velocity for the planet; when the planet moves faster than the local speed of sound, a bow shock forms in front of the planet, appreciably changing the character of the interaction between the gaseous atmospheric envelope and the stellar-wind gas.

Observations of hot Jupiters using the Hubble Space Telescope~\citep{Vidal-Madjar-2003, Ben-Jaffel-2007} indicated that Ly$\alpha$ absorption during the transit of the exoplanet HD\,209458\,b reached 9--15\%, while the decrease in the stellar brightness due to occultation of the planetary disk was only 1.8\%. A similar effect was also observed later in the C, O, and Si lines~\citep{Vidal-Madjar-2004, Ben-Jaffel-2010, Linsky-2010}. These observations led to the conclusion that the planet was surrounded by an extensive gaseous envelope. Similar results were obtained for the planets HD\,189733\,b and WASP-12\,b. Moreover, a transit observation of WASP-12\,b using the Hubble Space Telescope in 2009 revealed that the transit ingresses and egresses in different spectral ranges did not coincide, clearly indicating the presence of dense matter ahead of the planet at a distance of four or five planetary radii~\citep{Fossati-2010}.

Explaining available and planned observations of hot Jupiters requires a clear understanding of which physical phenomena predominate in the binary systems analyzed. A gas-dynamical model for investigating the interaction between the stellar wind and exoplanetary atmospheres was proposed in~\citep{Bisikalo-2013a, Bisikalo-2013b}. The results of calculations indicated that the solutions for typical hot-Jupiters moving in orbits with supersonic velocities are affected appreciably by the bow shock that is formed in front of the atmosphere. In particular, it was shown in~\citep{Bisikalo-2013b} that the type of gaseous envelope around the exoplanet depends on the position of the frontal collision point (FCP), where the dynamical wind pressure is equal to the atmospheric pressure of the exoplanet, resulting in the separation of matter flows incident on the atmosphere relative to the Roche lobe boundaries. Planets whose FCPs lie inside their Roche lobes have almost spherical envelopes characteristic of a classical atmosphere, only slightly distorted due to influence of the star and the interaction with the stellar-wind gas. When the FCP is beyond the Roche lobe of the planet, matter begins to outflow from the planetary atmosphere through the vicinities of the Lagrangian points \Lp1 and \Lp2, giving rise to a strongly asymmetric envelope. This type of object can also be divided into two types. If the dynamical pressure of the stellar-wind gas is high enough to stop the more powerful outflow through the inner Lagrangian point \Lp1, a quasi-closed stationary envelope with a complicated shape is formed in the binary system, as was first indicated in~\citep{Bisikalo-2013a}. If the wind cannot stop the flow through \Lp1, an open envelope is formed.

To determine the atmospheric properties of hot Jupiters, it is necessary to accurately determine the mass-loss rates for these different types of envelopes. This will enable us not only to estimate the characteristic lifetime of the gaseous envelope of a particular exoplanet, but also to impose specific constraints upon possible observational appearances of the envelopes that form. We have perfored numerical calculations for all three types of hot-Jupiter envelopes to determine the mass-loss rates from their atmospheres ($\dot{M}$). We have also analyzed the gas-dynamical properties of the flow structure arising due to the atmospheric mass loss of hot Jupiters. Section 2 briefly describes the numerical model, Section 3 provides the results of our numerical calculations, and Section 4 summarizes the main conclusions of this work.

\section{The Model}
\label{2}

As in~\citep{Bisikalo-2013b}, only gas-dynamical processes were considered when modeling the interaction between the incident flow of the stellar wind and the atmosphere of a hot Jupiter. The total effect of other physical phenomena in these systems is assumed to be much smaller than the effects due to the outflow from the planetary atmosphere beyond the planet's Roche lobe and the interaction between the planetary envelope and the stellar-wind gas.

The flow is described using a 3D system of equations for the gravitational gas dynamics, closed by the equation of state for an ideal, neutral monatomic gas (see, e.g,~\citep{Zhilkin-2012, Bisikalo-2013c}). The potential forces acting in the system were defined using the Roche potential, modified (in a way similar to~\citep{Mitsumoto-2005}) to take into account the acceleration of the stellar wind by introducing the parameter $\Gamma$, which is equal to zero in regions filled with stellar wind and unity in all other regions:

\begin{equation} \label{eq1}
\begin{split}
\Phi = 
- \Gamma \,
\bigg( 
\dfrac{G M_*   }{ \sqrt{x^2 + y^2 + z^2}} +  
\dfrac{G M_{pl}}{ \sqrt{(x-A)^2 + y^2 + z^2}} 
\bigg) 
- 
\dfrac{1}{2} \, \Omega^2 
\bigg[(x - A \dfrac{M_{pl}}{M_* + M_{pl}})^2  + y^2\bigg]
\end{split}
\end{equation}

where $G$ is the gravitational constant, $M_*$ the mass of the star, \Mpl\ the mass of the planet, $\Omega$ the rotational angular velocity of the binary system, and $A$ the semi-major axis of the planetary orbit. The origin of the coordinate system is at the center of the star, the $X$ axis is directed along the line connecting the centers of the binary components, the $Z$ axis is perpendicular to the orbital plane, and the $Y$ axis is oriented to make a right-handed coordinate system.

A Roe--Osher total variation diminishing (TVD) scheme with Einfeldt’s entropy correction was used for the calculations (see, e.g,~\citep{Zhilkin-2012, Bisikalo-2013c}). This is an explicit, higher-order approximation scheme that has a small numerical viscosity in regions with smooth solutions and does not lead to scattering of shocks. A gas-dynamical code parallelized over space was used. The calculations were performed at the Joint Supercomputer Center of the Russian Academy of Sciences using 324 processors.

The computations were performed on a non-uniform rectangular grid of $468 \times 468 \times 178$ cells for the $X$, $Y$, and $Z$ axes, respectively; the size of the computational domain is $(40 \times 40 \times 10)$\Rpl, where \Rpl\ is the photometric radius of the planet. A uniform space partition was specified along the $Z$ axis (perpendicular to the orbital plane), with the cells having a step of $\Delta z = 0.06$\Rpl. The grid was condensed along the $X$ and $Y$ axes toward the center of the exoplanet so that the cell size at the photometric radius of the planet ($\Delta x$, $\Delta y$ $ \simeq 0.04$\Rpl) was smaller than the minimum height scale for all the computed atmosphere models. In the process, the characteristic scale of the main flow elements at the boundaries of the computational domain (at the distance from the planet $r \simeq 20$\Rpl) was appreciably larger than the maximum cell size ($\Delta x$, $\Delta y$ $\leq 0.25$\Rpl).

To determine the mass-loss rate from the planetary atmosphere, surfaces were specified in the computational domain to cover all the flow elements: the outflows from \Lp1 and \Lp2 and the vortical wake that arises behind the planet. The mass flows obtained at cell margins when solving the numerical scheme were integrated over these surfaces. Next, the unperturbed stellar-wind mass flow passing through these surfaces was subtracted from the total mass loss to eliminate the effect of the stellar wind on the final estimates and to correctly determine the mass lost by the atmosphere.

\section{Results}
\label{3}

The atmospheric parameters for the typical hot Jupiter HD\,209458\,b were specified, according to the latest estimates obtained for this planet~\citep{Koskinen-2013}. We chose four pairs of parameters for the calculations (the temperature and number density of the atmospheric material at the photospheric radius) to cover the entire domain of their estimated range. The model parameters are presented in the Table~1. Note that, according to~\citep{Bisikalo-2013b}, this set of parameters enables us to obtain all three types of possible atmospheres.

\begin{table}[t!]
\caption{Parameters of the model atmospere: temperature and
number density of matter at the photometric radius}\label{tabular:models}
\begin{center}
\begin{tabular}{c|c|c}
\hline
Model number & $T$, K & $n, 10^{10} cm^{-3}$ \\
\hline
\hline
1 & 6000 & 2 \\
2 & 7000 & 5 \\
3 & 7500 & 10 \\
4 & 8000 & 20 \\
\hline
\end{tabular}

\end{center}
\end{table}

We used a simple model of the stellar wind, in which the wind velocity did not change in the neighborhood of the planet and the drop in the wind density was proportional to the inverse square of the distance to the star. Since the wind parameters for the solar-type star HD\,209458 are not known, these are assumed to be equal to the solar-wind parameters at a given distance from the star: $T_w=7.3 \times 10^5\,\mathrm{K}$, $n_w\sim10^4\,\mathrm{cm}^{-3}$, and $v_w\,=\,100\,\mathrm{km\,s^{-1}}$~\citep{Bisikalo-2013b, Withbroe-1988}, where $T_w$ is the wind temperature, $n_w$ the particle number density in the wind, and $v_w$ the radial wind velocity. The wind was assumed to have only a radial velocity in the fixed coordinate system.

As the planet moves with a supersonic velocity through the gas of the stellar wind, a bow shock and contact discontinuity arise in front of the planet, separating the wind material that has passed through the shock front from the static atmospheric material. A rarefaction zone with a strongly vortical flow is formed at the leeward side of the planet. In the supersonic flows considered, the solutions clearly split into different parts depending on the temperature. For all the models, the wind temperature appreciably exceeds the atmospheric temperature of the planet, suggesting that the maximum temperature (in the entire computational domain) will occur just behind the shock front, while the minimum temperature will be reached in the atmosphere and the rarefaction wave. Hence, these flow elements can be distinctly separated after temperature isosurfaces have been constructed in the regions of the shock and the contact discontinuity.

\subsection{Closed Atmosphere (Model~1)}
\label{3}


\begin{figure}[t!]
\begin{center}
\centering\epsfig{width=10cm,file=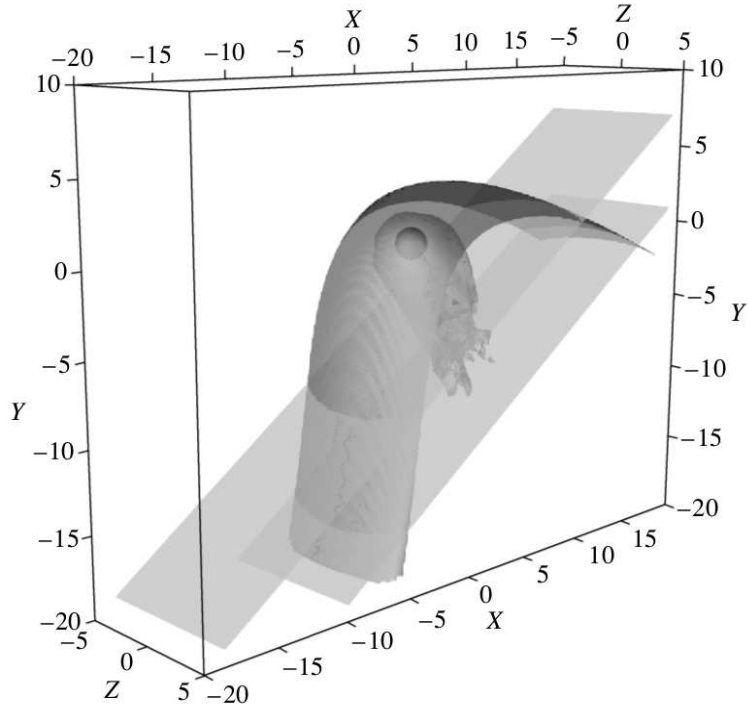}
\end{center}
\caption{Temperature isosurfaces corresponding to the shock and the contact discontinuity for Model~1. The planet is shown as a dark ball, and is located at the coordinate origin. All the coordinates are given in the units of \Rpl. The planes for which the intensities of the mass flows from the binary system are shown in Fig.~2 are also depicted.}\label{fig:temperature_6000}
\end{figure}


\begin{figure}[t!]
\begin{center}
\centering\epsfig{width=14cm,file=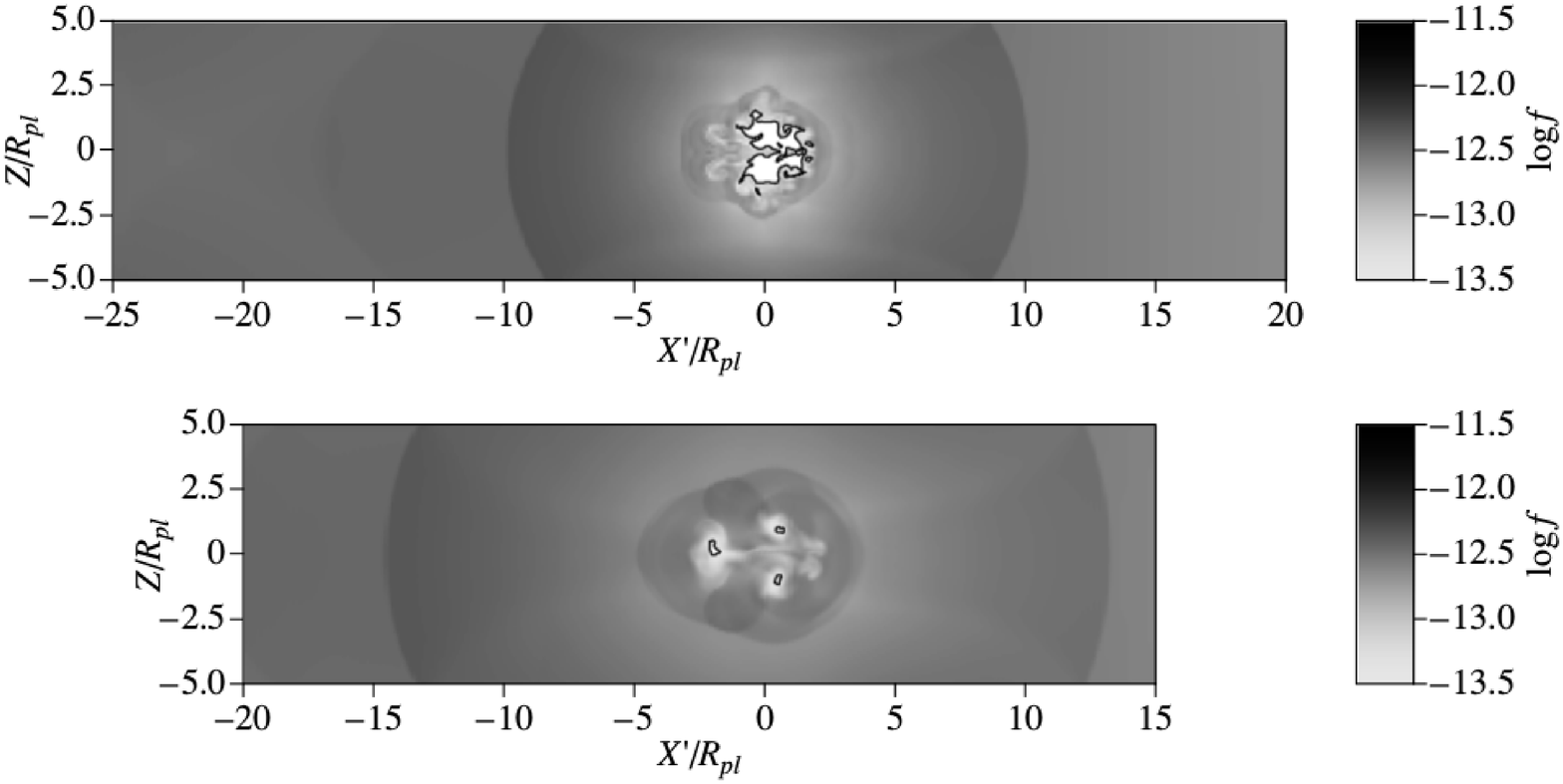}
\end{center}
\caption{Intensities of the mass flows through the planes depicted in Fig.~1. A logarithmic scale is used. The domains where the flow intensities are negative (matter moves toward the planet) or are lower than $3 \times 10^{-14}$\,\gscm\ are outlined by black contours and shown in white. The vertical axis corresponds to the $Z$ axis of the computational domain. The projection of the barycenter of the planet onto these planes is the point with coordinates (0, 0). The values on the horizontal axis increase in the direction that corresponds to the growth of $X$ and $Y$ in the computational domain. All the distances are indicated in the units of \Rpl.}\label{fig:flows_6000}
\end{figure}

Figure~1 shows the temperature isosurfaces corresponding to the shock and the contact discontinuity for Model~1 (the gaseous envelope is completely closed). The shock is symmetric (relative to the velocity vector of the incident flow); its shape is spherical near the FCP and becomes a Mach cone that diverges with distance from the planet. The surface of the contact discontinuity at the windward side of the planet is almost spherical.

Figure~2 shows the logarithmic mass-flow intensities through the planes depicted in Fig.~1. The first and second planes are at distances of $r_1 = 5$\Rpl\ and $r_2 = 10$\Rpl\ from the planet. The gray scale shows the intensities of flows (directed outward from the planet) higher than $f \geq 3 \times 10^{-14}$\,\gscm\ (or $f \geq 0.2 f_w$, where $f_w = \rho_w v_w = 1.6 \times 10^{-13}$\,\gscm\ is the intensity of the unperturbed stellar-wind mass flow at the distance of the semi-major axis from the star; this parameter changes only slightly within the computational domain). Solution regions where the flow intensities are negative or lower than the given value are outlined with black contours and filled in white. As can be seen in this figure, the planet has no clear outflows from the vicinities of the Lagrangian points; a vortical flow of matter that overall does not exceed the stellar-wind mass flow can be seen in the rarefied wake behind the planet. This solution indicates that weak mass loss from the atmosphere occurs according to the following scenario: some portion of the gas leaves the atmosphere in the direction of the leeward, rarefied region behind the planet, then slowly drifts away from the planet. In this case, the density of the matter flows from the atmosphere of the planet is lower than the wind flow density, implying that only an upper estimate for the mass-loss rate can be obtained. The total mass outflow in this model of a completely closed gaseous envelope is $\dot{M} \leq 10^{9}$\,\gs.

\subsection{Closed Atmosphere (Model~2)}
\label{3}


\begin{figure}[t!]
\begin{center}
\centering\epsfig{width=10cm,file=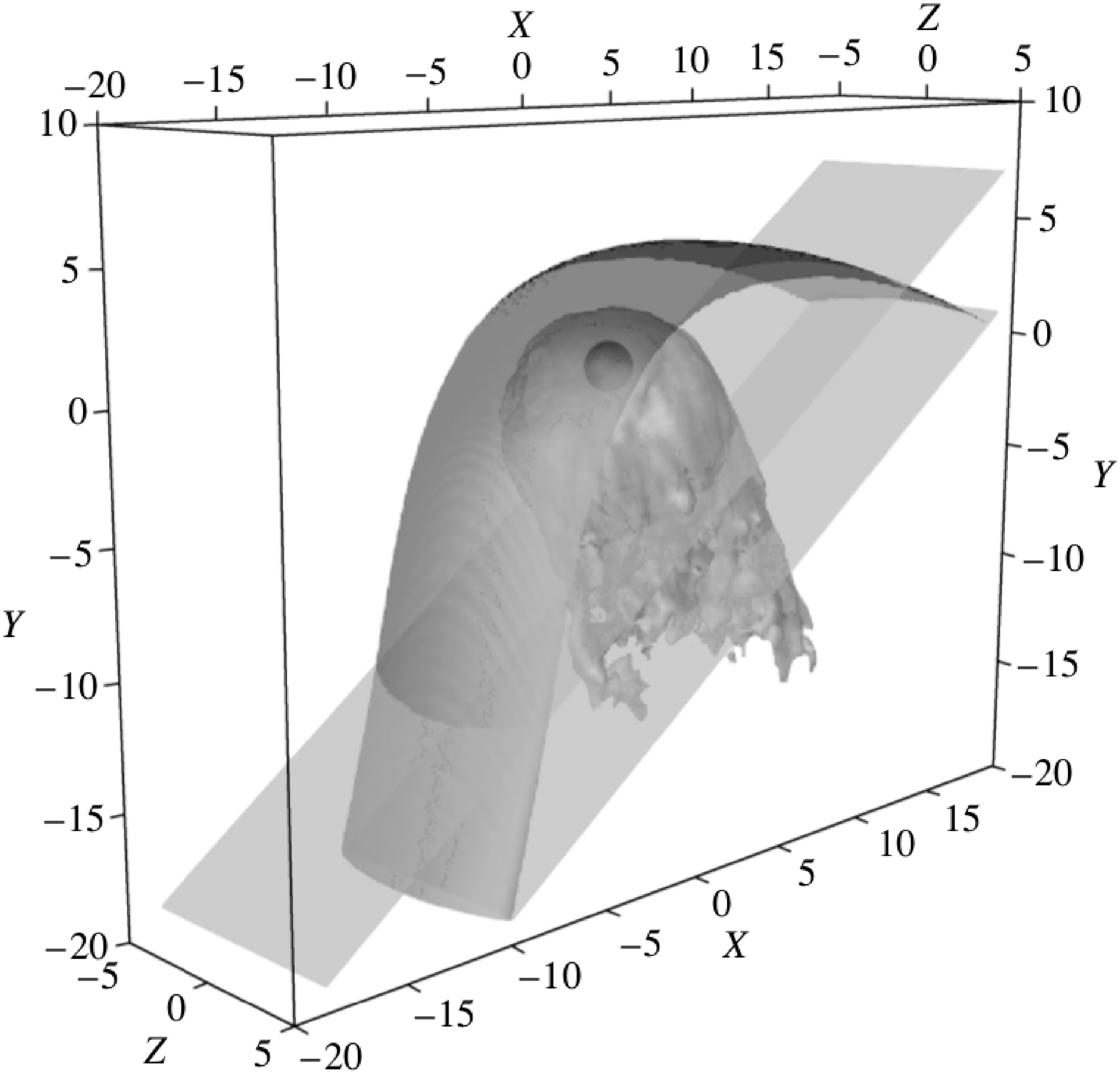}
\end{center}
\caption{Same as in Fig.~1 for Model~2.}\label{fig:temperature_7000}
\end{figure}


\begin{figure}[t!]
\begin{center}
\centering\epsfig{width=14cm,file=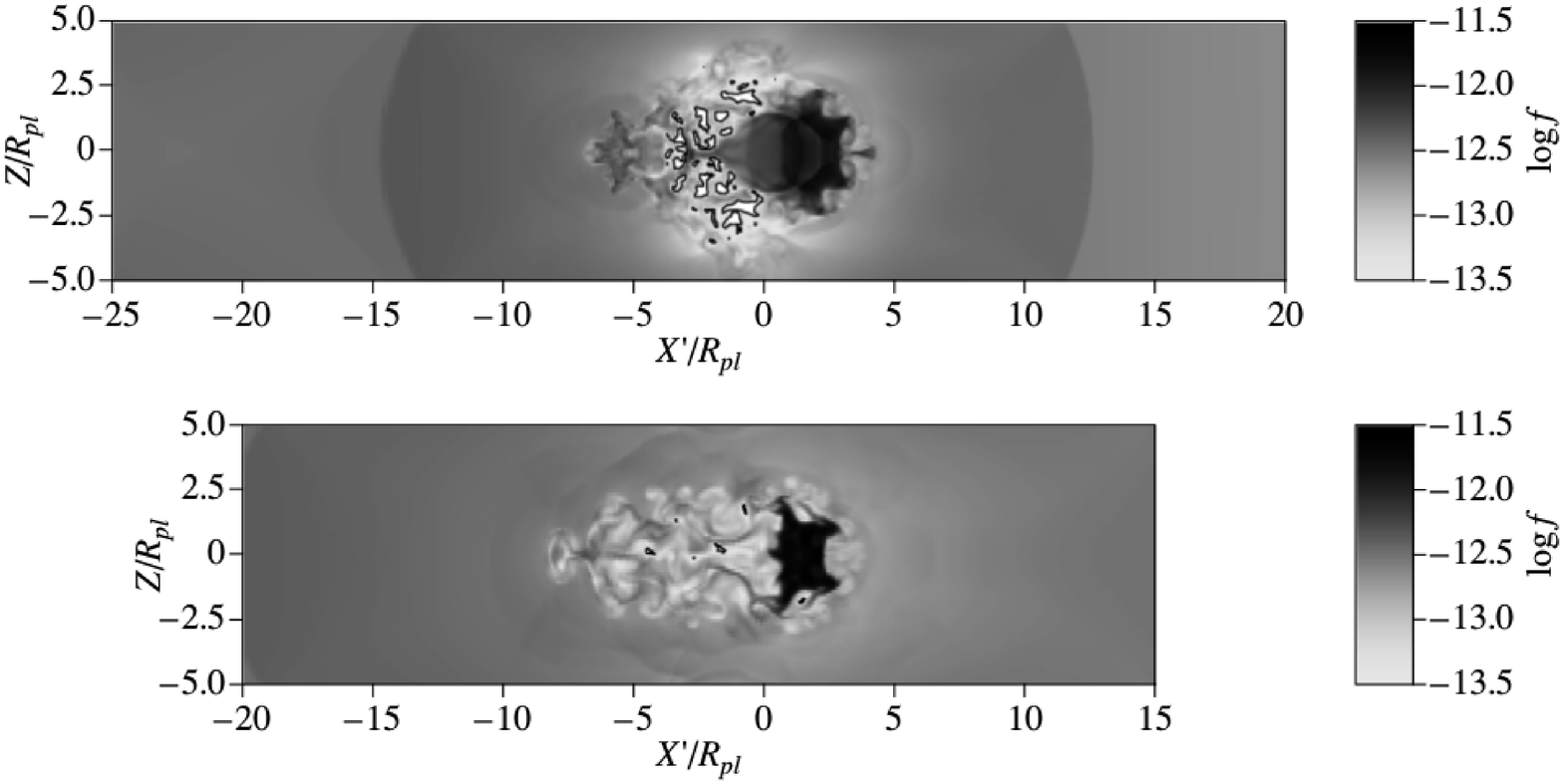}
\end{center}
\caption{Same as Fig.~2 for Model~2.}\label{fig:flows_7000}
\end{figure}

Figure~3 shows the main flow elements of Model~2. In contrast to the previous model, the atmosphere fills its Roche lobe almost completely, giving rise to a slight outflow from the vicinity of \Lp2. Since the Roche potential at \Lp1 is less than the potential at \Lp2, the outflow from \Lp1 would begin earlier in the absence of the stellar wind; however, the pressure of the stellar wind hinders the outflow in this binary system. Figure~3 supports this picture: the shape of the contact discontinuity is inflated compared to the previous model. The shock has a symmetrical, regular shape, similar to Model~1, but diverges further from the planet. The most protuberant part of the shock is beyond the planet’s Roche lobe.

Figure~4 shows the intensities of the mass flows through the planes depicted in Fig.~3 (all notation and parameters are the same as in Fig.~2). This figure indicates that there is a matter flow from the Lagrangian point \Lp2 and a rarefied vortical wake in this model. The area of the flow decreases with distance from the planet, while the mean density of the flow increases. The outer boundary of the outflow has a complex shape that seems to have arisen due to the development of instability of the contact discontinuity. The envelope in this model can be considered to be closed (although the flow from \Lp2 is partly open); the mass-loss rate is estimated to be $\dot{M} \simeq 10^{9}$\,\gs.

\subsection{Quasi-closed Atmosphere (Model~3)}
\label{3}


\begin{figure}[t!]
\begin{center}
\centering\epsfig{width=10cm,file=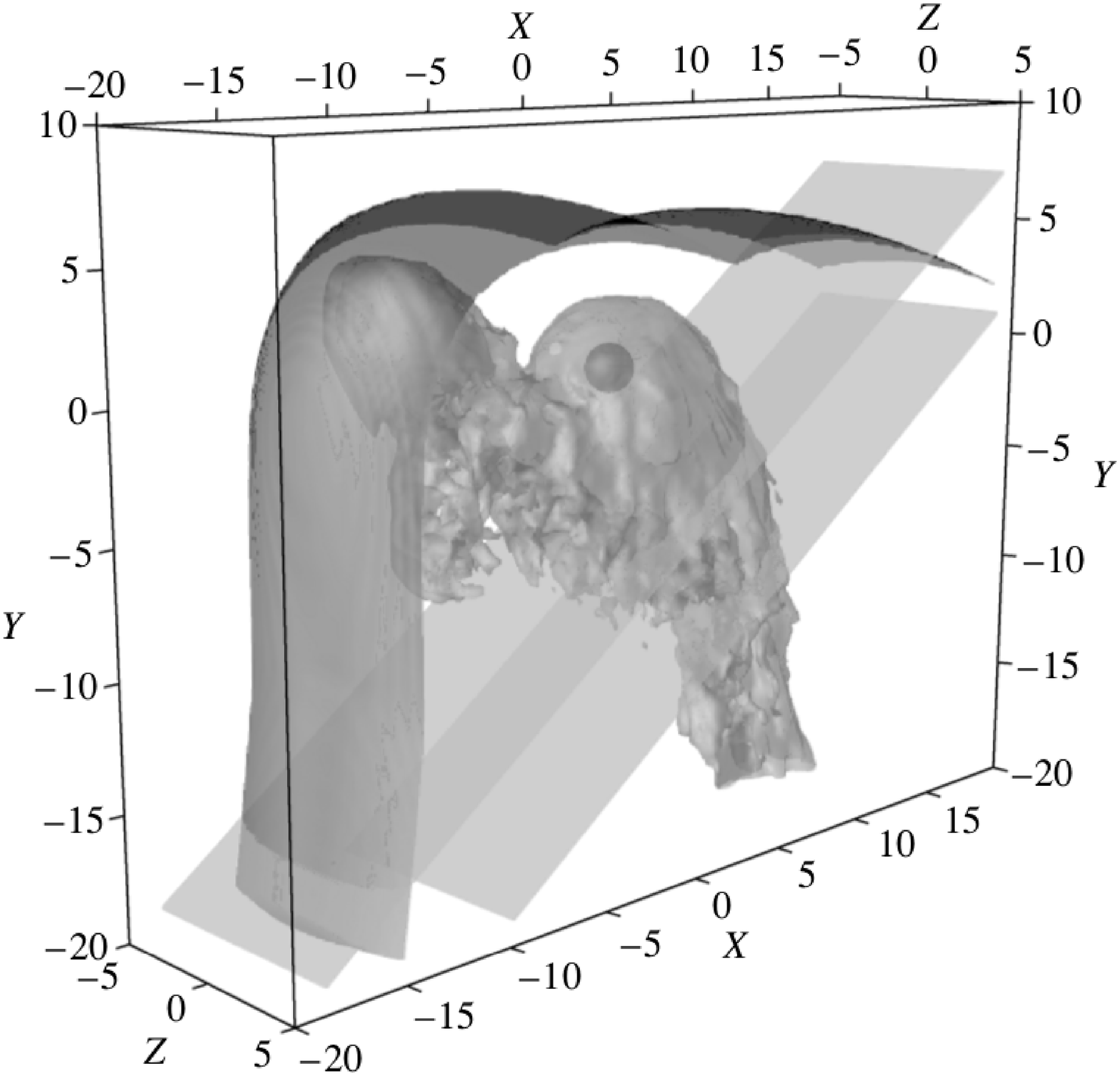}
\end{center}
\caption{Same as in Fig.~1 for Model~3.}\label{fig:temperature_7500}
\end{figure}


\begin{figure}[t!]
\begin{center}
\centering\epsfig{width=14cm,file=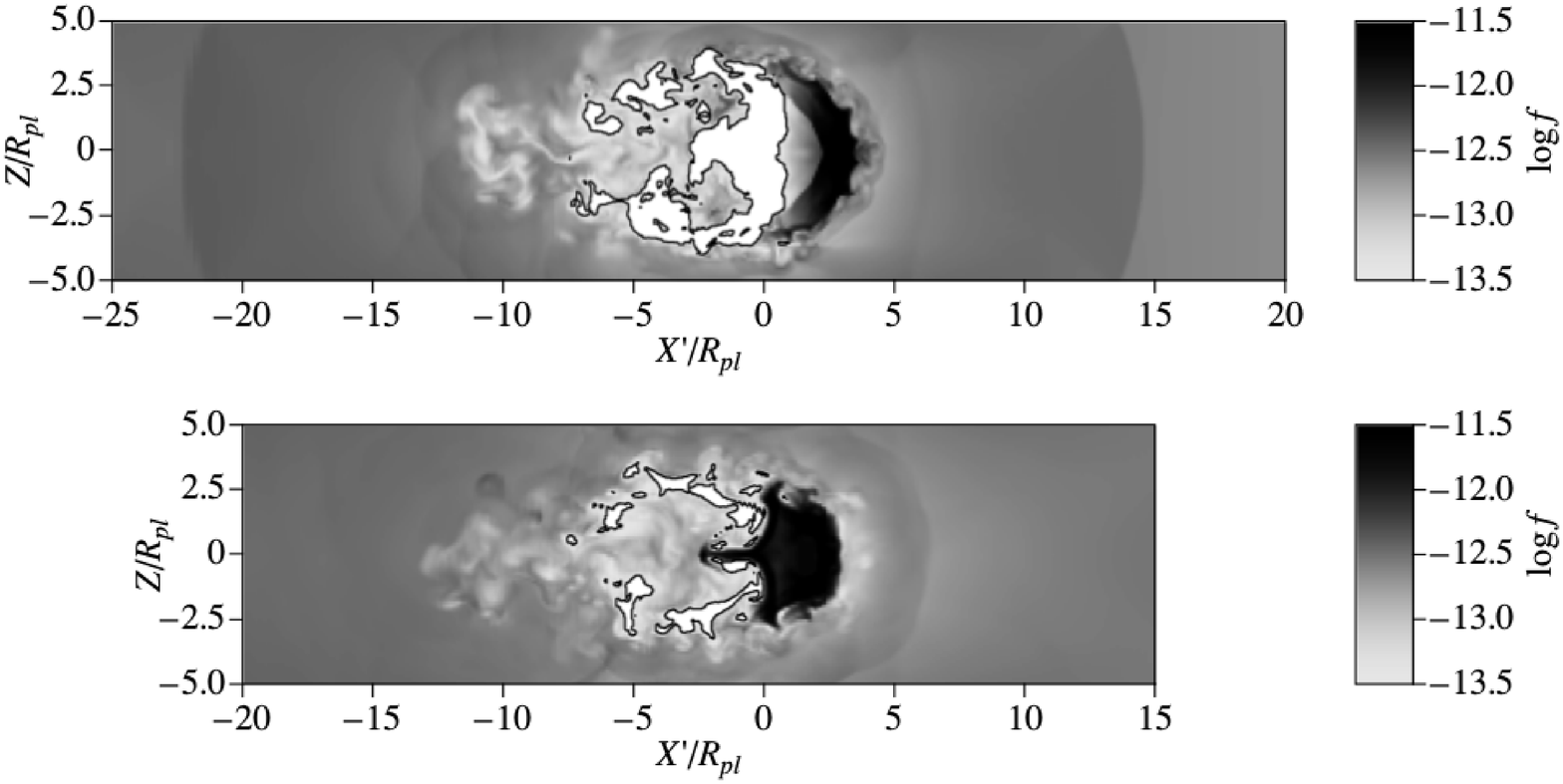}
\end{center}
\caption{Same as Fig.~2 for Model~3.}\label{fig:flows_7500}
\end{figure}

Figure~5 shows the main flow elements of Model~3. This figure indicates that there is a mass outflow through the Lagrangian point \Lp1 that is stopped by the dynamical pressure of the stellar wind at a distance of several \Rpl\ from \Lp1. In this case, the bow shock is a complex of two shocks, one formed just ahead of the planet and the other formed ahead of the outflow from \Lp1.

Figure~6 shows the intensities of mass flows through the planes depicted in Fig.~3; all the notation and parameters are the same as in Fig.~2. Since the outflow from \Lp1 is completely stopped by the stellar wind, the matter ejected through the vicinity of \Lp1 either returns back to the atmosphere or is shifted by the wind flow farther, in the direction of \Lp2. Hence, we conclude that the atmosphere loses mass mostly through the vicinity of \Lp2. Near the planet, the characteristic size of the flow from \Lp2 along the $Z$ direction exceeds its size along the $X'$ direction, and there is a region of reverse flow (the upper map in Fig.~4, corresponding to a distance of 5\Rpl\ from the planet).

The presence of the oppositely directed flow of matter suggests that the flow pattern in this region is fairly complex. The matter that leaves the atmosphere through the neighborhood of \Lp2 begins to expand freely, leading (since there is a rarefaction zone just after the planet) to a gas-pressure gradient; this, along with the gravitational force, shifts part of the flow backward toward the planet. Some of the matter becomes twisted due to the Coriolis force, and a stationary vortex is formed just behind the atmosphere, near \Lp2. The density of the vortical matter is several orders of magnitudes lower than the atmospheric density, although the shape of the Roche potential in this region allows the vortex to be in pressure balance with both the atmosphere and the flow from \Lp2. The vortex is comparatively small in size (several \Rpl), suggesting that the reverse flow disappears with distance from the planet, as can be seen in the lower map (Fig.~6). The density of the flow from \Lp2 is approximately equal to $f \simeq 10^{-11}$\,\gscm\ ($f \simeq 10^2 f_w$), which corresponds to the previous case; however, the cross-section area of the stream is appreciably larger in this case. Similarly to Model~2, instabilities develop at the outer boundary of the outflow, although with appreciably lower effect. This seems to be due to the greater density drop between the flow and the wind. For this model of a quasi-closed gaseous envelope, the mass-loss rate was estimated to be $\dot{M} \simeq 3 \times 10^{9}$\,\gs.


\begin{figure}[t]
\begin{center}
\centering\epsfig{width=10cm, file=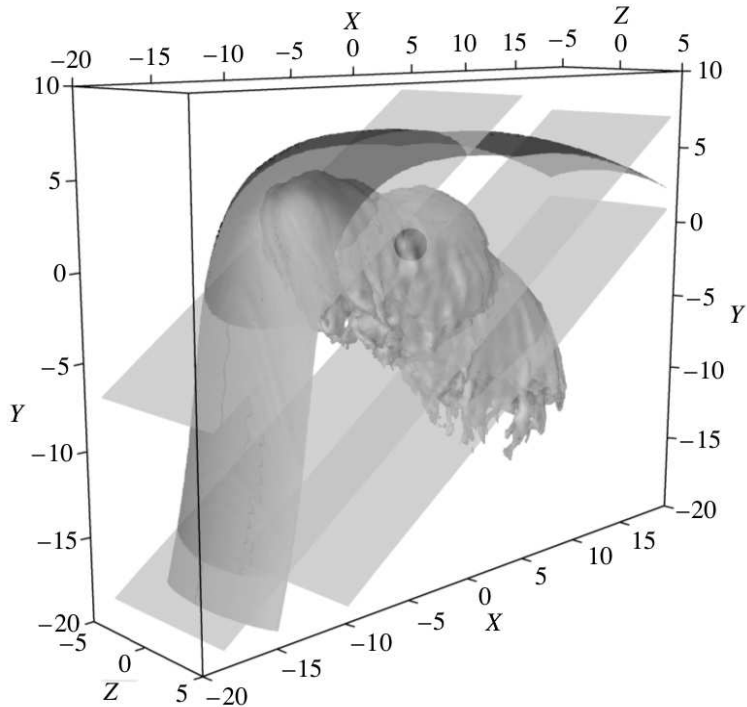}
\end{center}
\caption{Same as Fig.~1 for Model~4. The plane that crosses the flow from \Lp1 is also shown.}\label{fig:temperature_8000}
\end{figure}


\begin{figure}[t]
\begin{center}
\centering\epsfig{width=14cm,file=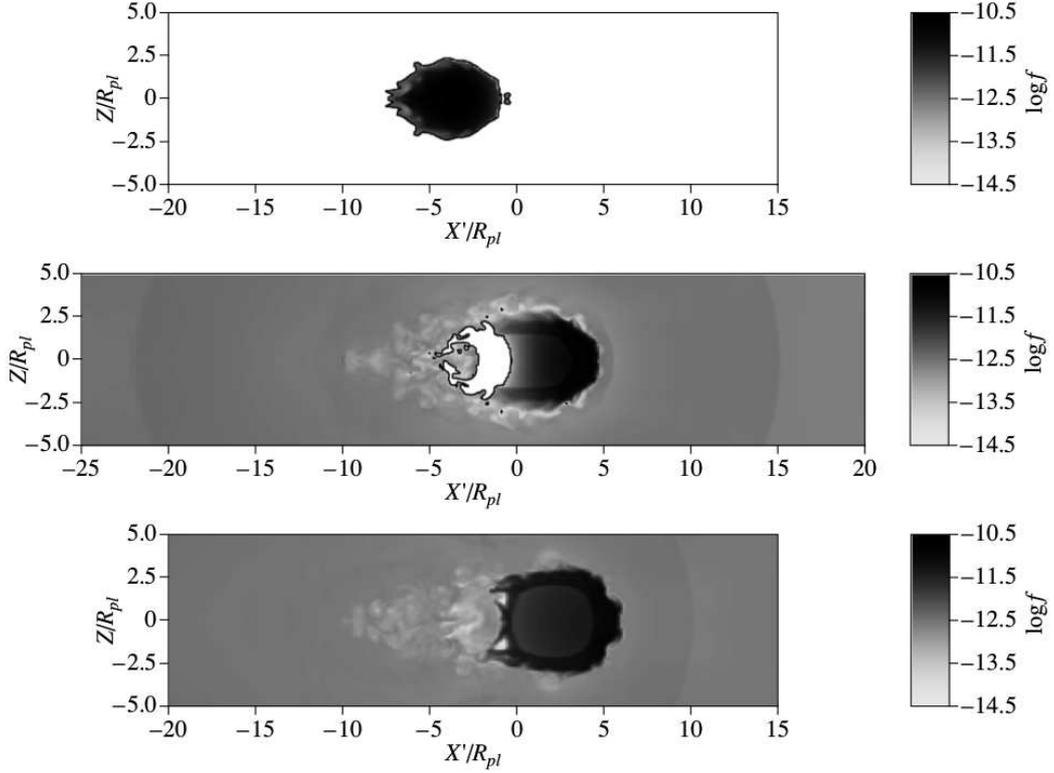}
\end{center}
\caption{Same as Fig.~2 for Model~4. The upper panel shows the flow distribution in a plane crossing the flow from \Lp1; flows directed outward from the planet are considered to be positive. Since the wind flows in this plane move toward the planet, the intensities of all the flows that are beyond the flow from \Lp1 are negative (these are located in the white region).}\label{fig:flows_8000}
\end{figure}

\subsection{Open Atmosphere (Model~4)}
\label{3}

Figure~7 shows the main flow elements of Model~4. In addition, three planes for which the intensities of matter flows from the binary system are shown in Fig.~8 are also depicted. These results indicate that two powerful outflows from the Lagrangian points \Lp1 and \Lp2 are formed in this solution; the flow from \Lp1 cannot be stopped by the dynamical pressure of the stellar wind. Importantly, this is an unstable binary system, and the figure presents the instantaneous condition of the system at a specific moment in time corresponding approximately to the same distance between the planet and the flow from \Lp1 as in Model~3. Subsequently, the flow from \Lp1 will continue to propagate and the matter will begin to leave the computational domain. As in Model~3, the bow shock is composed of a complex of two shocks, one in front of the planet and the other in front of the outflow from \Lp1.

Figure~8 shows the intensities of the mass flows across the three planes depicted in Fig.~7. In addition to the flows of matter from the binary system, which are similar to those considered in Models~1--3, this open-atmosphere solution provides an additional flow from the vicinity of \Lp1. This flow is calculated in a plane crossing the flow from \Lp1 and is shown in the upper map (Fig.~8). As can be seen in the upper and middle maps of Fig.~8, there is a constant flow of matter in the binary system from the Lagrangian points \Lp1 and \Lp2. The areas of the flows from \Lp1 and \Lp2 are almost the same, 20\Rpl$^2$, and their mean flow densities are $f \simeq 10^{-11}$\,\gscm\ ($f \simeq 10^2 f_w$). Analysis of the lower and middle maps (Fig.~8) indicates that, when there is an outflow from \Lp2, a zone where the flow is negative (it moves toward the planet) also forms near the planet, apparently for the same reasons as in Model~3. Note that the atmosphere in this model loses mass mainly from the neighborhood of the inner Lagrangian point: the mass losses from \Lp1 and \Lp2 are $\dot{M}_{L_1} \simeq 17 \times 10^{9}$\,\gs\ and $\dot{M}_{L_2} \simeq 14 \times 10^{9}$\,\gs, respectively. The resulting mass-loss rate from the atmosphere in this case is $\dot{M} \simeq 3 \times 10^{10}$\,\gs.

\section{Conclusions}
\label{4}

We have estimated the mass-loss rates for three possible types of envelopes around hot Jupiters~\citep{Bisikalo-2013b}: closed, quasi-closed, and open. 3D gas-dynamical modeling of the interaction between a typical hot-Jupiter with the parameters of HD\,209458\,b and the stellar wind was performed, and four models corresponding to the latest estimates of the atmospheric parameters for this planet were considered~\citep{Koskinen-2013}. We chose the model parameters so as to obtain all three types of atmospheres, while remaining within the range of known estimates.

The solutions obtained for Models~1 and 2 correspond to systems with closed envelopes; there is a small outflow of matter from the vicinity of \Lp2 in Model~2. The upper limit for the mass-loss rates from these atmospheres is $10^9$\,\gs. The envelope becomes quasi-closed in Model~3; however, the outflow from the neighborhood of the inner Lagrangian point \Lp1 is stopped by the stellar wind and does not propagate farther than several radii \Rpl, and matter is lost mainly through the vicinity of \Lp2 on the leeward side. The mass-loss rate in Model~3 is $3 \times 10^9$\,\gs, only slightly higher than the corresponding values for the closed envelopes. In Model~4, the atmosphere becomes open and most of the matter is lost due to outflow through \Lp1, which cannot be stopped by the stellar wind. According to our estimates, the mass-loss rate for the open atmosphere is $3\times10^{10}$\,\gs. The estimates of mass-loss rates obtained for the closed and quasi-closed enevelopes correspond well to values derived from observations of HD\,209458\,b \citep{Shematovich-2010, Murray-Clay-2009, Koskinen-2010, Garcia Munoz-2007}.

Note that the solutions we have obtained enable us to conclude with confidence that large non-spherical envelopes can exist around hot Jupiters. Regardless of the fact that the sizes of these envelopes appreciably exceed the Roche lobe of the planet, the mass-loss rates can remain low enough to ensure a stationary atmosphere and a long lifetime for the planet. This possibility for hot Jupiters to have stationary, aspherical envelopes fundamentally changes widespread approaches used to analyze observational data for exoplanetary atmospheres.

\section*{Acknowledgments}
\label{5}

This work was supported by the Basic Research Program of the Presidium of the Russian Academy of Sciences, the Russian Foundation for Basic Research (projects 12-02-00047 and 13-02-00077), and the Program of Support to Leading Scientific Schools of the Russian Federation (NSh-3620.2014.2)

\end{document}